\documentclass[12pt,a4paper]{article}
\usepackage{times,mathptm,amsmath,epsfig}
\newcommand{\blin}{\textsf{BAYES-LIN}}
\newcommand{\ls}{\textsf{LISP-STAT}}
\title{\blin: An object-oriented environment for Bayes linear
local computation}
\author{Darren J.~Wilkinson\thanks{%
Department of Statistics, University of Newcastle, Newcastle upon Tyne, NE1 7RU,
ENGLAND. Email: \texttt{d.j.wilkinson@ncl.ac.uk} WWW:
\texttt{http://www.ncl.ac.uk/{}\~{}ndjw1/} 
}
\\
Copyright \copyright\ 1997
} 
\date{\today}

\newcommand{\bd}{\ensuremath{[B/D]}}

\newcommand{\var}[2][]{\mathrm{Var}_{#1}\left(#2\right)}
\newcommand{\cov}[3][]{\mathrm{Cov}_{#1}\left(#2,#3\right)}

\newcommand{\ex}[2][]{\mathrm{E}_{#1}\left(#2\right)}

\newcommand{\trans}{\mbox{}^\mathrm{T}}
\newcommand{\inv}{\mbox{}^{-1}}

\newcommand{\be}{\begin{eqnarray}}
\newcommand{\ee}{\end{eqnarray}}
\newcommand{\beq}{\begin{equation}}
\newcommand{\eeq}{\end{equation}}

\begin{document}
\maketitle

\noindent\textsf{The latest version of the \blin\ software and documentation
(including the latest version of this document), can be obtained from
the \blin\ WWW page:}
\begin{verbatim}
               http://www.ncl.ac.uk/~ndjw1/bayeslin/
\end{verbatim}
\begin{abstract}
\blin\ is an extension of the \ls\ object-oriented statistical
computing environment, which adds to \ls\ some object prototypes
appropriate for carrying out local computation \emph{via}
message-passing between clique-tree nodes of Bayes linear belief
networks. Currently the \blin\ system represents a rather low-level
set of tools for a back-end computational engine, together with
diagnostic graphics for understanding the effects of adjustments on
the moral graph. A GUI front end,
allowing interactive formulation of DAG models could be easily added,
but is currently missing from the system. This document provides a
very brief introduction to the system, by means of a work-through of
two example computations, followed by a list of variables, functions,
objects and methods provided by the system. 
\end{abstract}

\newpage
\tableofcontents 
\newpage

\section{Introduction}
\subsection{Bayes linear methods}
Bayes linear methods are a form of Bayesian statistics, which
acknowledge the difficulties associated with the full modelling,
specification, and conditioning required by distributional Bayesian
statistics, and instead try to make best possible use of partial
specifications, based on means, variances and
covariances. Unsurprisingly, much of the theory is formally identical
to inference in multivariate Gaussian Bayesian networks, but
interpretation of results is generally different. This document
assumes a working knowledge
of the basic tools of the Bayes linear methodology. An introduction to
Bayes linear methods is given in \cite{fgcross}. An introduction to
(non-local) computational issues can be found in \cite{gwblincomp}.
The foundations of the theory are discussed in \cite{mgpriorinf},
\cite{mgrevexch}, and \cite{mgrevprev}. On-line, an introduction to
the theory can be found in \cite{blmadj}, from the Bayes Linear
Methods WWW home page: \verb$http://fourier.dur.ac.uk:8000/stats/bayeslin/$

\subsection{\ls}
\ls\ is an interpreted, object-oriented environment for statistical
computing, described in \cite{lispstat}. This document assumes a
working knowledge of \ls, and the basics of object-oriented
programming. On-line, \ls\ information is available from the \ls\ WWW
home page:
\verb$http://www.stat.umn.edu/~luke/xls/xlsinfo/xlsinfo.html$

\subsection{Local computation}
\blin\ carries out local computation \emph{via} message-passing between
adjacent nodes of a clique-tree representing the statistical model of
interest. Again, local computation in Bayesian networks is a huge area,
and this document assumes a working knowledge of graphical models,
conditional independence and some of the ideas behind local
computation. The best introduction to all of these areas is
\cite{pearlbook}. In particular, Chapter 3 of that volume deals with
all of the relevant graph-theoretic concepts, and Section 7.2 gives an
introduction to graphical Gaussian models. 

\subsection{Installing and running \blin}
You need a working \ls\ system installed before you attempt to install
\blin. The following instructions are for a UNIX system with an
\textsf{XLISP-STAT} installation, but
installing on other systems should be similar.
Note that the graphics work best on systems with at least a 16 bit colour
display. If you only have an 8 bit display (256 colours), make sure
that most are free for use by \blin. The graphics will not work on
displays poorer than 8 bit colour.
Create a new directory for the \blin\ system. Download the \blin\
software from the \blin\ WWW page:
\verb$http://www.ncl.ac.uk/~ndjw1/bayeslin/$
and put into the new directory. In
this new directory type:
\begin{verbatim}
% gunzip blin01a.tar.gz
% tar -xvf blin01a.tar
% gzip blin01a.tar
\end{verbatim}
You should then be able to run \ls\ with the \blin\ extensions simply by
running
\begin{verbatim}
% xlispstat
\end{verbatim}
from within this directory. You can check that the extensions are
loaded by typing in some of the following commands in the \ls\
listener window.
\begin{verbatim}
> (help 'create-tree-node)
> (send moral-node-proto :help)
> (send tree-node-proto :help :observe)
\end{verbatim}
In general, to make sure the extensions are loaded, use the expression
\begin{verbatim}
> (require "bayeslin")
\end{verbatim}
When you are satisfied that the extensions are loaded, exit \blin.
\begin{verbatim}
> (exit)
\end{verbatim}
In order to run the examples, simply call \ls\ with the example as
first argument. \emph{eg.}
\begin{verbatim}
% xlispstat ex-dlm
\end{verbatim}
or
\begin{verbatim}
% xlispstat ex-mdlm
\end{verbatim}
These two examples will be explained in the following sections.

\section{A ``toy'' dynamic linear model}
\subsection{Description of the model}
The \blin\ code for this example can be found in the file
\verb$ex-dlm.lsp$, which is part of the standard \blin\ distribution.
The example concerns a very simple model for 3 observations in time. 
The model can be written in the form of a locally constant DLM.
\begin{align*}
X_t &= \theta_t + \nu_t \\
\theta_t &= \theta_{t-1} + \omega_t 
\end{align*}
$X_t$ denotes the observation at time $t$ ($t=1,2,3$), which is dependent on the
\emph{state} of the system at time $t$, $\theta_t$. The variables
$\nu_t$ and $\omega_t$ are incidental noise terms. The model is
initialised by specifying beliefs about the initial state of the
system; in this case, $\ex{\theta_1}=1,\ \var{\theta_1}=1$, and the
variance of the noise terms; in this case,
$\var{\omega_t}=\var{\nu_t}=1$. Of course, for such a simple model, a
non-local analysis is trivial, since the expectation vector for the
entire system, and the variance matrix for the entire system can be
written down and worked with
directly.
$\ex{\theta_1,\theta_2,\theta_3,X_1,X_2,X_3}=(1,1,1,1,1,1)\trans$,
\[
\var{\theta_1,\theta_2,\theta_3,X_1,X_2,X_3}=\left(\begin{array}{cccccc}
1&1&1&1&1&1\\
1&2&2&1&2&2\\
1&2&3&1&2&3\\
1&1&1&2&1&1\\
1&2&2&1&3&2\\
1&2&3&1&2&4
\end{array}\right)
\]
However, for the example in the next section, such explicit non-local
analysis will not be possible.

\subsection{Graphical models}
 The graph for this model is shown
below.
\vspace{0.5cm}

\noindent
\epsfig{file=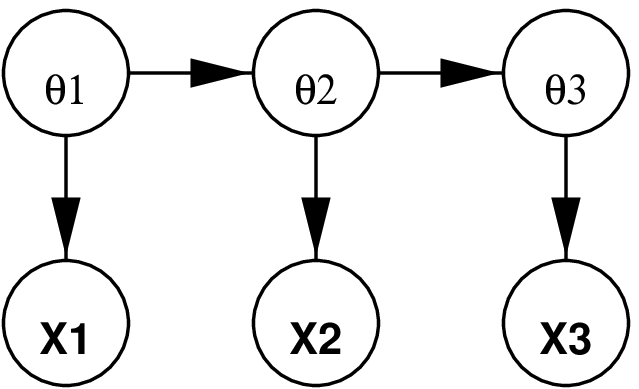,height=2in}
\vspace{0.5cm}

Since there are no unmarried parents, this graph can be moralised
simply by dropping arrows.
\vspace{0.5cm}

\noindent
\epsfig{file=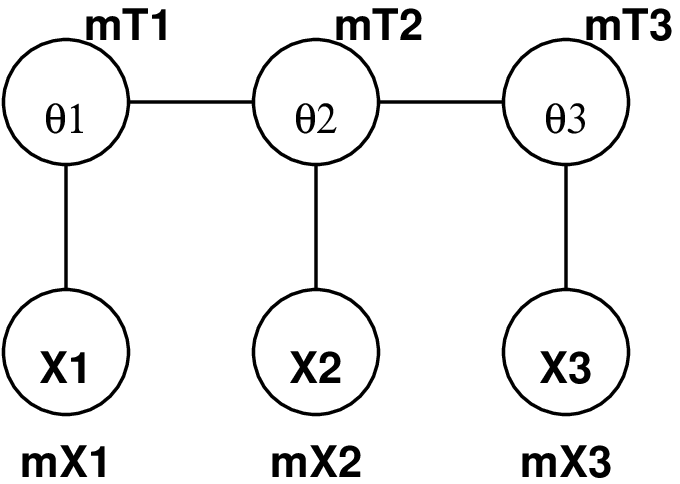,height=2in}
\vspace{0.5cm}

Now, since there are no cycles, this graph is already triangulated, so
the clique tree may be formed as follows.
\vspace{0.5cm}

\noindent
\epsfig{file=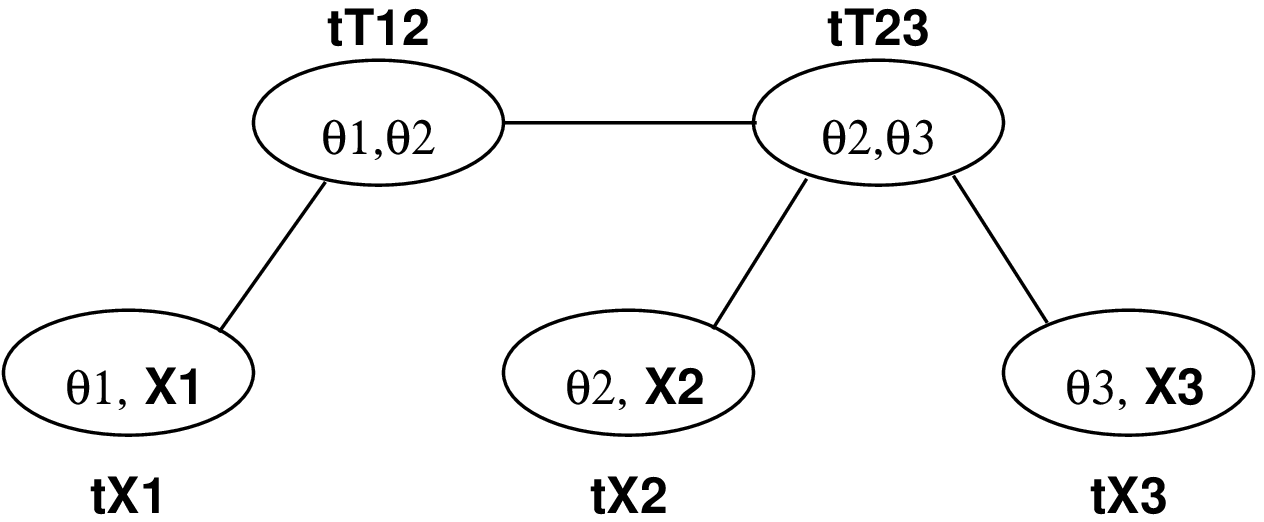,height=2in}
\vspace{0.5cm}

\subsection{Defining the clique-tree}
\blin\ carries out computation on the clique-tree, and displays
results on the moral graph. Therefore, both need to be introduced to
the \blin\ system. Since all information and computations are carried
out on the clique-tree (in fact, computations can be carried out
without defining a moral graph at all), this is defined
first. Appropriate code for defining the tree nodes is shown below.
{\small
\begin{verbatim}
(create-tree-node 'tX1 '(t1 x1) #(1 1) #2a((1 1) (1 2)) '(tT12))
(create-tree-node 'tX2 '(t2 x2) #(1 1) #2a((2 2) (2 3)) '(tT23))
(create-tree-node 'tX3 '(t3 x3) #(1 1) #2a((3 3) (3 4)) '(tT23))
(create-tree-node 'tT12 '(t1 t2) #(1 1) #2a((1 1) (1 2)) '(tX1 tT23))
(create-tree-node 'tT23 '(t2 t3) #(1 1) #2a((2 2) (2 3)) '(tT12 tX2 tX3))
\end{verbatim}}
The global function \verb$create-tree-node$ is used to define each
node in turn. The function expects five arguments. The first argument
is a symbol to point to the resulting tree-node object. The second is
a list of variables which the node contains. The third and fourth are
the expectation vector and variance matrix for the variable list, and
the fifth is a list of neighbouring tree nodes.
 Next, the moral nodes
are defined.

\subsection{Defining the moral graph}
{\small
\begin{verbatim}
(create-moral-node 'mX1 '(x1) 'tX1 "mX1" '(mT1))
(create-moral-node 'mX2 '(x2) 'tX2 "mX2" '(mT2))
(create-moral-node 'mX3 '(x3) 'tX3 "mX3" '(mT3))
(create-moral-node 'mT1 '(t1) 'tX1 "mT1" '(mX1 mT2))
(create-moral-node 'mT2 '(t2) 'tX2 "mT2" '(mX2 mT1 mT3))
(create-moral-node 'mT3 '(t3) 'tX3 "mT3" '(mX3 mT2))
\end{verbatim}}
The global function \verb$create-moral-node$ is used to define each
node in turn. The first is a symbol to bind the object to. The second
is a variable list. The third is a clique-tree node which contains all
of the variables at this node (such a node always exists). The fourth
is a string to be used for plotting purposes, and the fifth is a list
of neighbouring moral graph nodes. Next, some plotting positions are
defined by sending a \verb$:location$ message to each moral node
object. 
\begin{verbatim}
(send mX1 :location '(0.2 0.8))
(send mX2 :location '(0.5 0.8))
(send mX3 :location '(0.8 0.8))
(send mT1 :location '(0.2 0.2))
(send mT2 :location '(0.5 0.2))
(send mT3 :location '(0.8 0.2))
\end{verbatim}
This step may be omitted if one is not interested in plotting of
results. The locations are on a $(0,1)$ scale for $x$ and $y$
coordinates, respectively. The origin is the top-left of the plot
window. The model is now completely specified. Before carrying out
adjustment, we create plot windows to show diagnostic information.

\subsection{Adjustments}
\begin{verbatim}
(create-moral-plot 'myplot)
(create-global-moral-plot 'myplot2)
\end{verbatim}
This creates a plot window with the name \verb$myplot$ to show partial
adjustment information, and another, \verb$myplot2$, to show global
adjustment information. Note that high-quality colour Encapsulated PostScript
output is produced for each plot after each redraw of the screen, and
stored in the files \verb$mpw.eps$ and \verb$gmpw.eps$ respectively.

We are now in a position to carry out adjustments. Suppose that
variable $X_1$ is observed to be $x_1=3$. This information can be
introduced into the graph by sending the following message to the
appropriate moral graph node.
\begin{verbatim}
(send mx1 :observe '(x1) #(3))
\end{verbatim}
In general, one can observe a list of variables, provided all
variables are contained in the moral graph node receiving the
message. The message is passed on to the appropriate clique-tree node,
and then propagated around the clique-tree. We can tell our plot
object to gather information from the tree for display, as follows.
\begin{verbatim}
(send myplot :record)
\end{verbatim}
The plot should now show how information flows around the moral graph
(more on this later). Note that although information has been
introduced into the graph, it has not been \emph{absorbed} into it,
and that further information can not be introduced until it has. This
can be understood by sending some messages to the graph, and looking
at the return values. If the expectation and variance of the first
observable node is examined
\begin{verbatim}
(send mx1 :ex)
(send mx1 :var)
\end{verbatim}
it can be seen that it retains its \emph{a priori} values. However,
one can also ask for \emph{adjusted} expectations and variances.
\begin{verbatim}
(send mx1 :aex)
(send mx1 :avar)
\end{verbatim}
Similar queries can be sent to the third moral graph node.
\begin{verbatim}
(send mx3 :ex)
(send mx3 :var)
(send mx3 :aex)
(send mx3 :avar)
\end{verbatim}
When we are finished examining the effects of the current adjustment,
and wish to add further information into the graph, the current
information should be absorbed.
\begin{verbatim}
(send mx1 :absorb)
\end{verbatim}
The absorbing makes the adjusted information the new prior
information, ready for the next adjustment. This can be verified by
looking at the new expectation and variance for the third observation.
\begin{verbatim}
(send mx3 :ex)
(send mx3 :var)
\end{verbatim}
Finally, we can introduce new information, record it, and then absorb
it, before examining the results.
\begin{verbatim}
(send mx2 :observe '(x2) #(-1))
(send myplot :record)
(send mx2 :absorb)

(send mx3 :var)
(send mx3 :ex)
\end{verbatim}
The two plot windows should now look similar to the following.

\noindent\epsfig{file=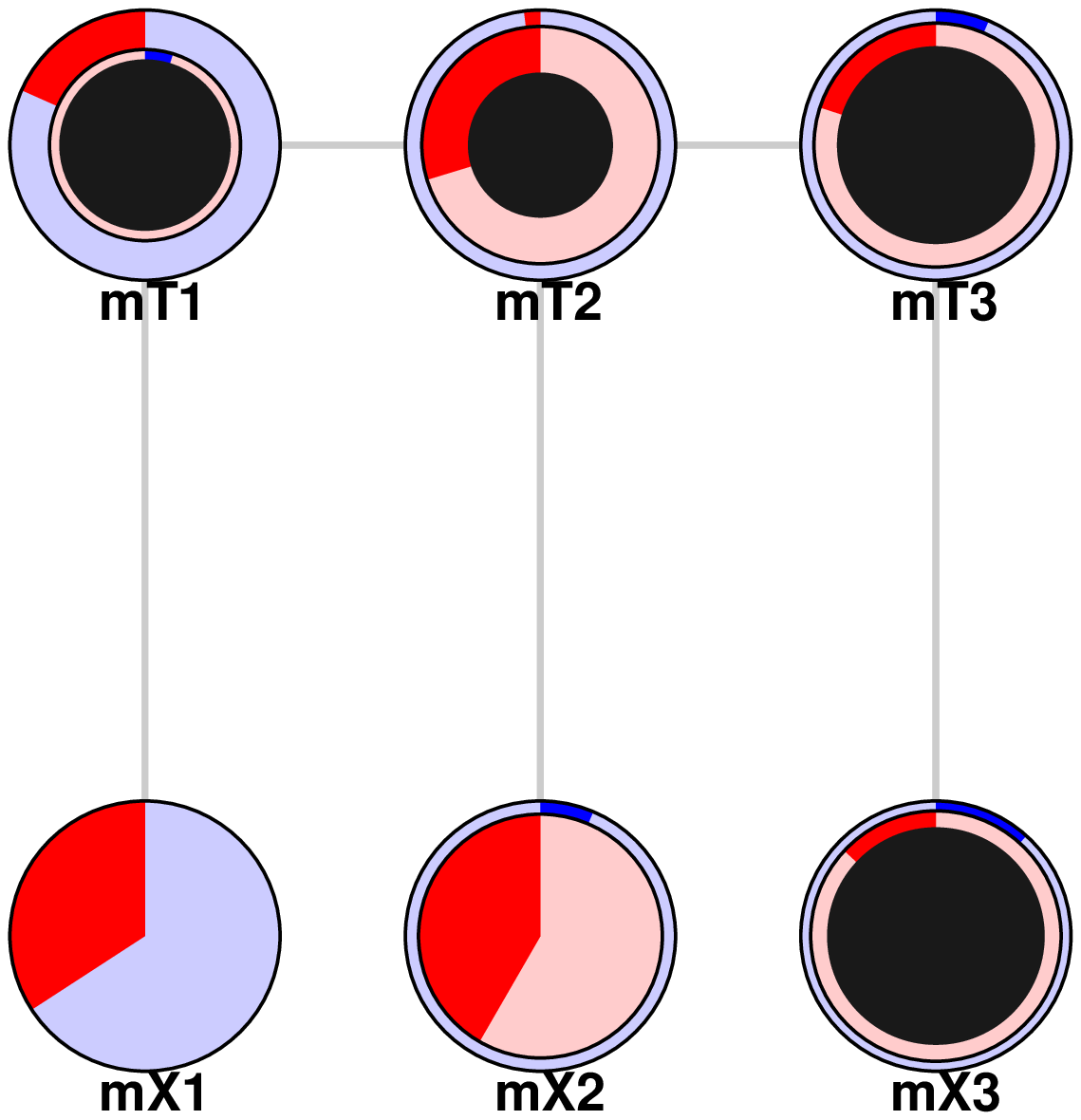,width=4in}

\noindent\epsfig{file=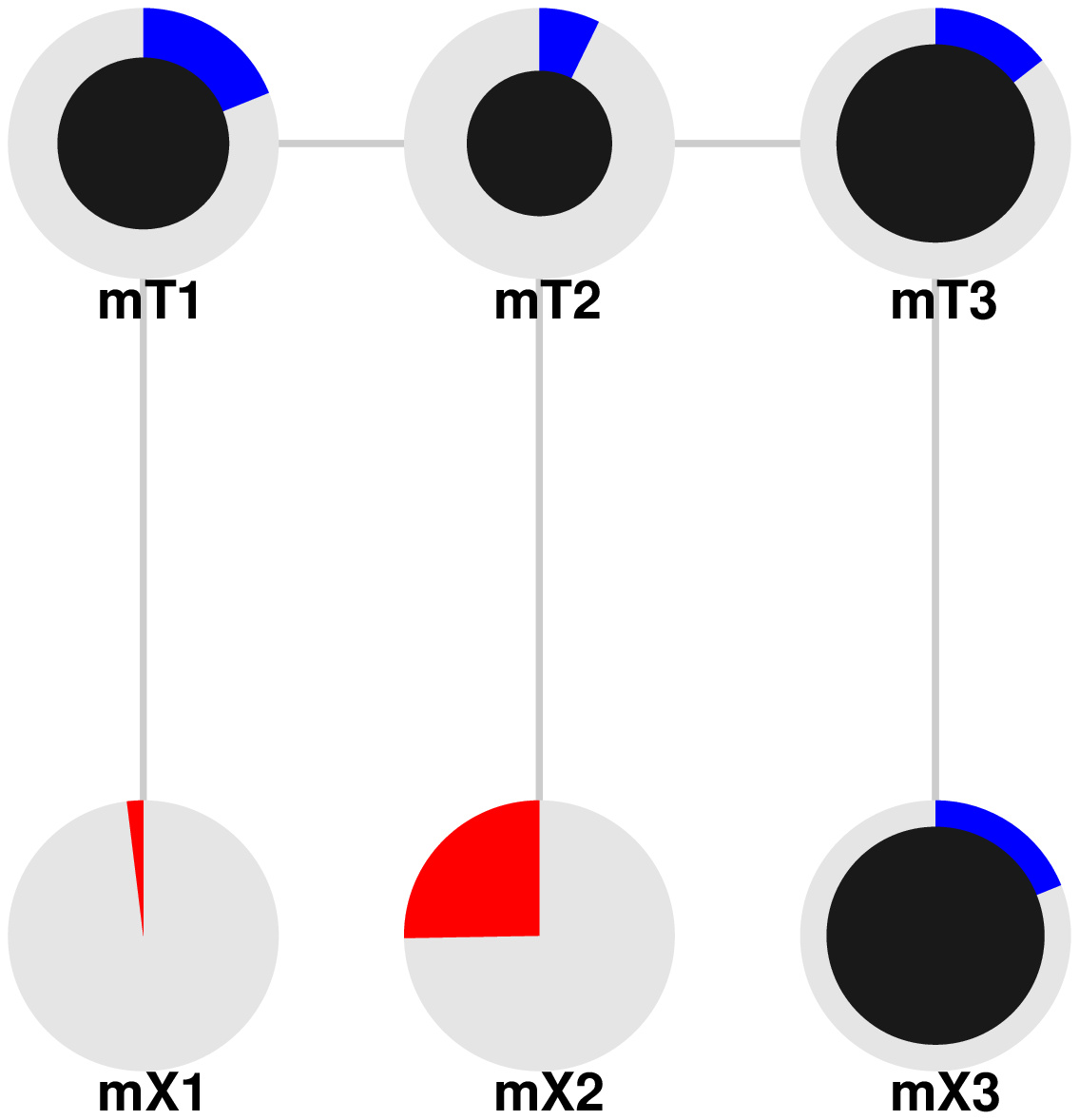,width=4in}

\subsection{Interpreting the graphics window}
Whenever an observation is made and recorded, a portion is removed
from the outside of each node. The area removed is proportional to the
variance resolved by the adjustment. Consequently, the radius removed
is proportional to the standard deviation resolved. Therefore, when a
node is fully observed, there is no dark centre left remaining. For
other nodes, the size of the dark centre is proportional to the
proportion of original uncertainty left remaining. For multivariate
nodes, the Bayes linear concept of \emph{resolution} is used.

The additional red and blue shadings give an indication of the changes in
expectation, relative to \emph{a priori} uncertainty. Red shadings
indicate changes larger than expected, and blue shadings represent
changes in expectation smaller than expected%
\footnote{Note that changes in expectation smaller than expected can
still be of concern, since they are indicative of a possible
under-utilisation of prior information.}%
. The amount of red and blue shading increases as the ``degree of
surprisingness'' increases. The amount of shading is a transformation
of the Bayes linear concept of \emph{size-ratio}. The transformation
can be user-specified by redefining the plot-object's \verb$:sr-map$
method appropriately.

\section{Computation for a large multivariate DLM} 
\subsection{Description of the model}
The \blin\ code for this example can be found in the file
\verb$ex-mdlm.lsp$, which is part of the standard \blin\
distribution.
The following data represent weekly sales of six soft drinks packs
from a wholesale depot. 
\begin{verbatim}
  51 27 1  4  6 3
 113 55 0  7 15 4
 103 71 0 10 16 7
   .  . .  .  . .
   .  . .  .  . .
\end{verbatim}
Clearly a multivariate time series model is
required for such data. The following multivariate locally constant
DLM is adopted.
\begin{align*}
X_t &= \theta_t + \nu_t \\
\theta_t &= \theta_{t-1} + \omega_t 
\end{align*}
This is the same model as used in the last example, but here all of
the variables denote random vectors of dimension six. The model is
specified in the following way. There are 35 observations, and so $t$
runs from $1$ to $35$ for the actual observations. However, for this
model, it was felt more convenient to 
initialise the model at $t=0$. The initial state of the system was
specified as $\ex{\theta_0}=(50,50,50,50,50,50)\trans$ and
$\var{\theta_0}=diag(900,900,900,900,900,900)$. The covariance
structure for the noise terms was specified to be
$\var{\nu_t}=V,\ \var{\omega_t}=W$, where
\begin{align*}
V=&
\left(\begin{array}{rrrrrr}
2420.36&387.33&20.39&165.27&44.56&58.61\\
387.33&263.85&3.85&71.51&23.48&3.27\\
20.39&3.85&30.79&3.58&1.26&5.99\\
165.27&71.51&3.58&139.72&23.12&11.33\\
44.56&23.48&1.26&23.12&50.01&4.78\\
58.61&3.27&5.99&11.33&4.78&44.21
\end{array}\right)
\\ W=&
\left(\begin{array}{rrrrrr}
1112.49&272.47&22.52&66.45&31.56&27.84\\
272.47&195.50&11.53&30.07&18.51&15.37\\
22.52&11.53&29.64&5.54&4.67&6.28\\
66.45&30.07&5.54&78.91&14.04&8.03\\
31.56&18.51&4.67&14.04&40.50&7.32\\
27.84&15.37&6.28&8.03&7.32&32.97
\end{array}\right)
\end{align*}
See \cite{djwgdlm} for an explanation of the given
specification. These specifications determine the model, but note that
there are $6\times 4\times 35 + 6=846$ variables in this problem
(assuming that we are interested in the noise terms). This problem is
about at the limit of the size which can be tackled by a
brute force approach, making a local computation approach particularly
attractive.

\subsection{Graphical models}
We are interested in making inferences about the noise
terms in this example (in order to help diagnose deficiencies of the
model), and so the noise terms need to be included in the model. The
first part of the
DAG for this structure is therefore as follows (note that the DAG
nodes are all multivariate).
\vspace{0.5cm}

\noindent
\epsfig{file=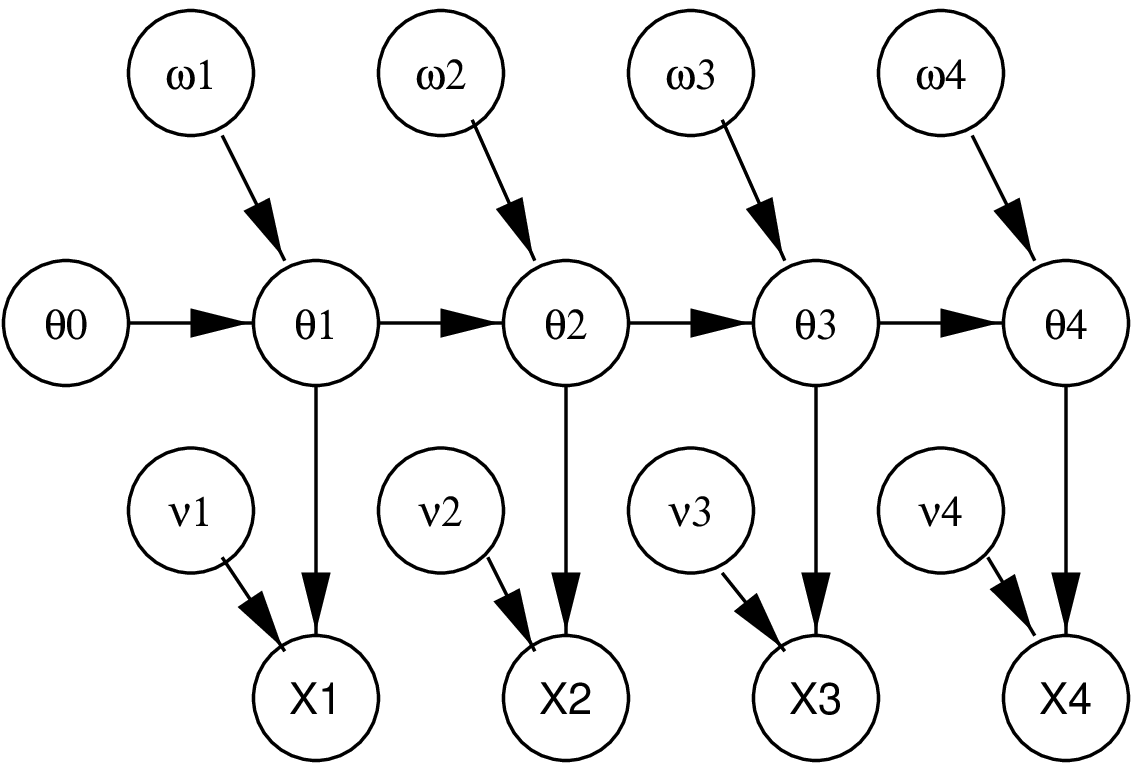,width=4in}
\vspace{0.5cm}

Marrying parents and dropping arrows gives the moral graph for the
problem (note that the moral graph nodes are all multivariate).
\vspace{0.5cm}

\noindent
\epsfig{file=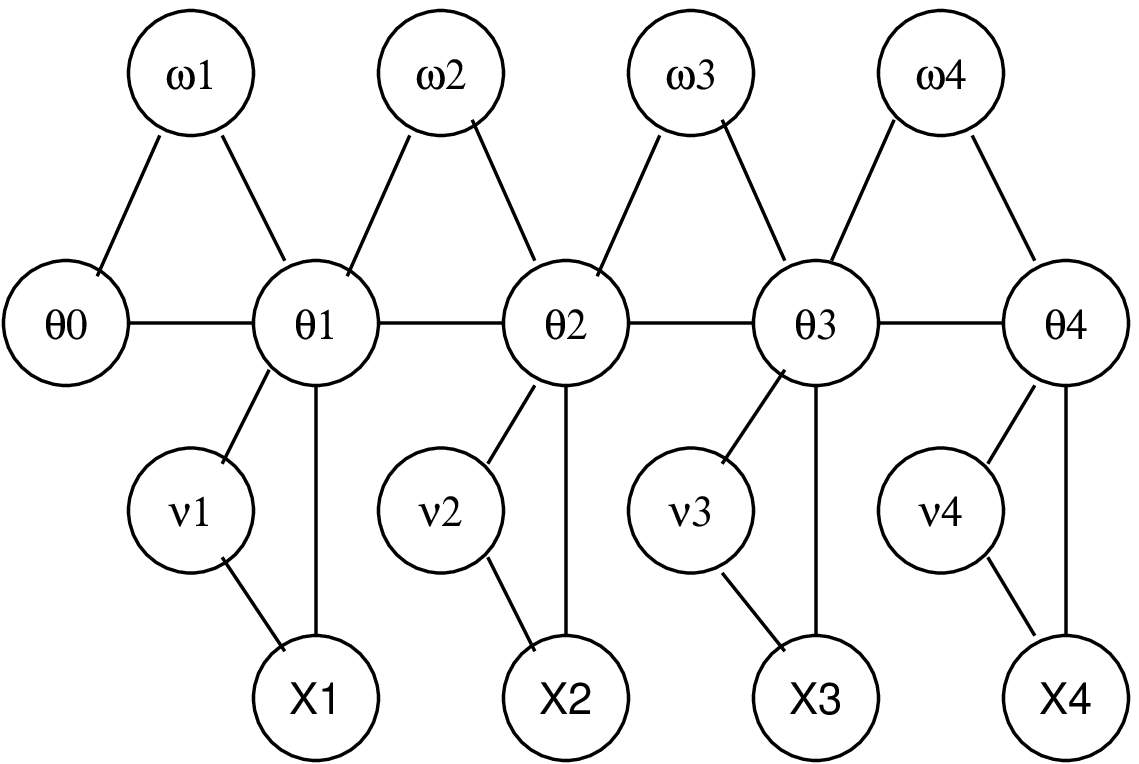,width=4in}
\vspace{0.5cm}

Again we are fortunate in the sense that the moral graph is
ready-triangulated, and so the clique-tree can be directly constructed
as follows.
\vspace{0.5cm}

\noindent
\epsfig{file=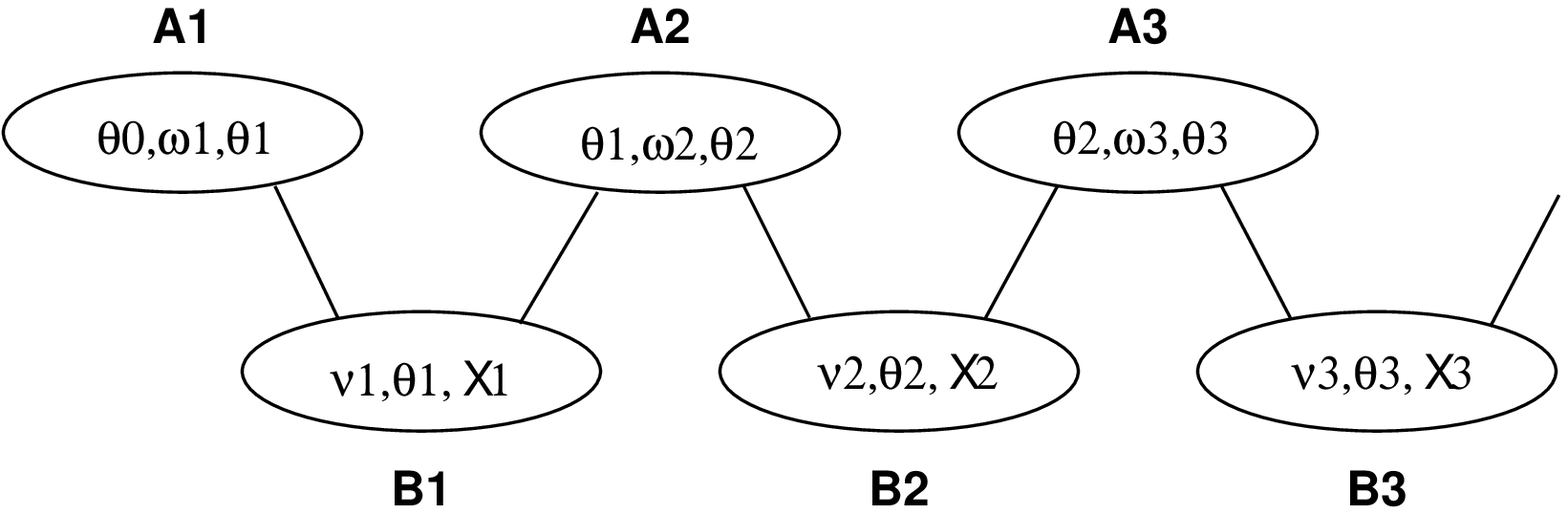,width=5in}
\vspace{0.5cm}

\subsection{Some constants}
The code for such a problem can be constructed as follows. First, the
data is read, turned into a matrix, and some constants are defined.
{\scriptsize
\begin{verbatim}
(def mydata (read-data-columns "ex-mdlm.dat"))
(def data (make-array '(6 35) :initial-contents mydata))

(def v
#2a(
    (  2420.36       387.330       20.3907       165.274       44.5645       58.6081    )
    (  387.330       263.850       3.85480       71.5054       23.4794       3.26543    )
    (  20.3907       3.85480       30.7910       3.58376       1.25836       5.98943    )
    (  165.274       71.5054       3.58376       139.715       23.1193       11.3268    )
    (  44.5645       23.4794       1.25836       23.1193       50.0087       4.78345    )
    (  58.6081       3.26543       5.98943       11.3268       4.78345       44.2135    )
   )
)

(def w
#2a(
    (  1112.49       272.473       22.5176       66.4472       31.5611       27.8440    )
    (  272.473       195.499       11.5298       30.0701       18.5065       15.3737    )
    (  22.5176       11.5298       29.6411       5.53985       4.66797       6.27591    )
    (  66.4472       30.0701       5.53985       78.9076       14.0421       8.03411    )
    (  31.5611       18.5065       4.66797       14.0421       40.5035       7.32038    )
    (  27.8440       15.3737       6.27591       8.03411       7.32038       32.9728    )
   )
)

(def e0 (coerce (repeat 50 6) 'array))
(def ee0 (coerce (append (repeat 50 12) (repeat 0 6)) 'array))
(def w0 (diagonal (repeat 900 6)))
(def zero66 (diagonal (repeat 0 6)))
\end{verbatim}}
These are all self-explanatory.

\subsection{Defining the clique-tree}
We can now create the type B cliques
as follows.
{\scriptsize
\begin{verbatim}
(dolist (i (iseq 1 35))
	 (create-tree-node
	  (intern (format nil "b~a" i))
	  (list (intern (format nil "x1.~a" i))
		(intern (format nil "x2.~a" i))
		(intern (format nil "x3.~a" i))
		(intern (format nil "x4.~a" i))
		(intern (format nil "x5.~a" i))
		(intern (format nil "x6.~a" i))
		(intern (format nil "theta1.~a" i))
		(intern (format nil "theta2.~a" i))
		(intern (format nil "theta3.~a" i))
		(intern (format nil "theta4.~a" i))
		(intern (format nil "theta5.~a" i))
		(intern (format nil "theta6.~a" i))
		(intern (format nil "nu1.~a" i))
		(intern (format nil "nu2.~a" i))
		(intern (format nil "nu3.~a" i))
		(intern (format nil "nu4.~a" i))
		(intern (format nil "nu5.~a" i))
		(intern (format nil "nu6.~a" i))
		)
	  ee0
	  (bind-rows (bind-columns
		      (+ w0 (* w i) v)
		      (+ w0 (* w i))
		      v)
		     (bind-columns
		      (+ w0 (* w i))
		      (+ w0 (* w i))
		      zero66)
		     (bind-columns
		      v
		      zero66
		      v)
		     )
	  (if (= i 35)
	      (list (intern (format nil "a~a" i)))
	    (list (intern (format nil "a~a" i))
		  (intern (format nil "a~a" (+ i 1))))
	    )
))
\end{verbatim}
}
Note that the expression \verb$(intern (format nil "b~a" i))$ means
``create the Lisp symbol \texttt{b}\textit{i}, where \textit{i} is a
variable''. This trick is used a lot for the construction of big
models with a repetitive structure. Next, the type A cliques can be
constructed, in a very similar way.
{\scriptsize
\begin{verbatim}
(dolist (i (iseq 1 35))
	 (create-tree-node
	  (intern (format nil "a~a" i))
	  (list (intern (format nil "theta1.~a" (- i 1)))
	        (intern (format nil "theta2.~a" (- i 1)))
	        (intern (format nil "theta3.~a" (- i 1)))
	        (intern (format nil "theta4.~a" (- i 1)))
	        (intern (format nil "theta5.~a" (- i 1)))
	        (intern (format nil "theta6.~a" (- i 1)))
	        (intern (format nil "theta1.~a" i))
	        (intern (format nil "theta2.~a" i))
	        (intern (format nil "theta3.~a" i))
	        (intern (format nil "theta4.~a" i))
	        (intern (format nil "theta5.~a" i))
	        (intern (format nil "theta6.~a" i))
		(intern (format nil "omega1.~a" i))
		(intern (format nil "omega2.~a" i))
		(intern (format nil "omega3.~a" i))
		(intern (format nil "omega4.~a" i))
		(intern (format nil "omega5.~a" i))
		(intern (format nil "omega6.~a" i))
		)
	  ee0
	  (bind-rows (bind-columns
		      (+ w0 (* w (- i 1)))
		      (+ w0 (* w (- i 1)))
		      zero66)
		     (bind-columns
		      (+ w0 (* w (- i 1)))
		      (+ w0 (* w i))
		      w)
		     (bind-columns
		      zero66
		      w
		      w)
		     )
	  (if (= i 1)
	      (list (intern (format nil "b~a" i)))
	    (list (intern (format nil "b~a" i))
		  (intern (format nil "b~a" (- i 1))))
	    )
))
\end{verbatim}}

\subsection{Defining the moral graph}
Next, moral graph nodes need to be created, for diagnostic plotting
purposes. Since there isn't room on the average computer screen for
the moral graph for all 35 time point, the structure will only be
constructed for the first 8 time points only.
{\scriptsize
\begin{verbatim}
;; number of moral nodes to create and plot
(def plotnum 8)
(create-moral-node (intern "theta.0")
		   (list (intern "theta1.0")
			 (intern "theta2.0")
			 (intern "theta3.0")
			 (intern "theta4.0")
			 (intern "theta5.0")
			 (intern "theta6.0"))
		   (intern "a1")
		   "Theta(0)"
		   (list (intern "theta.1")
			 (intern "omega.1"))
		   )
(send (symbol-value (intern "theta.0")) :location (list (/ 1 (+ plotnum 2)) 0.4))
(dolist (i (iseq 1 plotnum))
  ;; create the theta node
  (create-moral-node (intern (format nil "theta.~a" i))
		    (list (intern (format nil "theta1.~a" i))
			  (intern (format nil "theta2.~a" i))
			  (intern (format nil "theta3.~a" i))
			  (intern (format nil "theta4.~a" i))
			  (intern (format nil "theta5.~a" i))
			  (intern (format nil "theta6.~a" i)))
		    (intern (format nil "b~a" i))
		    (format nil "Theta(~a)" i)
		    (if (< i plotnum)
			(list (intern (format nil "omega.~a" i))
			      (intern (format nil "nu.~a" i))
			      (intern (format nil "x.~a" i))
			      (intern (format nil "theta.~a" (- i 1)))
			      (intern (format nil "theta.~a" (+ i 1)))
			      (intern (format nil "omega.~a" (+ i 1)))
			      )
		      (list (intern (format nil "omega.~a" i))
			    (intern (format nil "nu.~a" i))
			    (intern (format nil "x.~a" i))
			    (intern (format nil "theta.~a" (- i 1)))
			    )
		      )
		    )
  (send (symbol-value (intern (format nil "theta.~a" i))) :location
	(list (* (+ i 1) (/ 1 (+ plotnum 2))) 0.4)
	)
  ;; create the omega node
  (create-moral-node (intern (format nil "omega.~a" i))
		     (list (intern (format nil "omega1.~a" i))
			   (intern (format nil "omega2.~a" i))
			   (intern (format nil "omega3.~a" i))
			   (intern (format nil "omega4.~a" i))
			   (intern (format nil "omega5.~a" i))
			   (intern (format nil "omega6.~a" i)))
		     (intern (format nil "a~a" i))
		     (format nil "Omega(~a)" i)
		     (list (intern (format nil "theta.~a" i))
			   (intern (format nil "theta.~a" (- i 1))))
		     )
  (send (symbol-value (intern (format nil "omega.~a" i))) :location
	(list (* (+ i 0.5) (/ 1 (+ plotnum 2))) 0.2)
	)
  ;; create the nu node
  (create-moral-node (intern (format nil "nu.~a" i))
		     (list (intern (format nil "nu1.~a" i))
			   (intern (format nil "nu2.~a" i))
			   (intern (format nil "nu3.~a" i))
			   (intern (format nil "nu4.~a" i))
			   (intern (format nil "nu5.~a" i))
			   (intern (format nil "nu6.~a" i)))
		     (intern (format nil "b~a" i))
		     (format nil "Nu(~a)" i)
		     (list (intern (format nil "theta.~a" i))
			   (intern (format nil "x.~a" i)))
		     )
  (send (symbol-value (intern (format nil "nu.~a" i))) :location
	(list (* (+ i 0.5) (/ 1 (+ plotnum 2))) 0.6)
	)
  ;; create the x node
  (create-moral-node (intern (format nil "x.~a" i))
		     (list (intern (format nil "x1.~a" i))
			   (intern (format nil "x2.~a" i))
			   (intern (format nil "x3.~a" i))
			   (intern (format nil "x4.~a" i))
			   (intern (format nil "x5.~a" i))
			   (intern (format nil "x6.~a" i)))
		     (intern (format nil "b~a" i))
		     (format nil "X(~a)" i)
		     (list (intern (format nil "theta.~a" i))
			   (intern (format nil "nu.~a" i)))
		     )
  (send (symbol-value (intern (format nil "x.~a" i))) :location
	(list (* (+ i 1) (/ 1 (+ plotnum 2))) 0.8)
	)
)
\end{verbatim}}
The plots can now be created in the usual way.
\begin{verbatim}
(create-moral-plot 'myplot)
(create-global-moral-plot 'myplot2)
\end{verbatim}

\subsection{Adjustments}
The first 6 weeks of observations will be added into the model.
{\scriptsize
\begin{verbatim}
;; Sequentially introduce the data
(dolist (i (iseq 1 6))
(format t "~&Data for week ~a" i)
(send (symbol-value (intern (format nil "b~a" i))) :observe
      (list 
            (intern (format nil "x1.~a" i))
            (intern (format nil "x2.~a" i))
            (intern (format nil "x3.~a" i))
            (intern (format nil "x4.~a" i))
            (intern (format nil "x5.~a" i))
            (intern (format nil "x6.~a" i))
	    )
      (select (column-list data) (- i 1))
      )
(send myplot :record)
(send (symbol-value (intern (format nil "b~a" i))) :absorb)
)
\end{verbatim}}
The resulting plot windows give a good impression of the adjustment
process, and the way information flows forward and backwards through
time in such models.

\noindent\epsfig{file=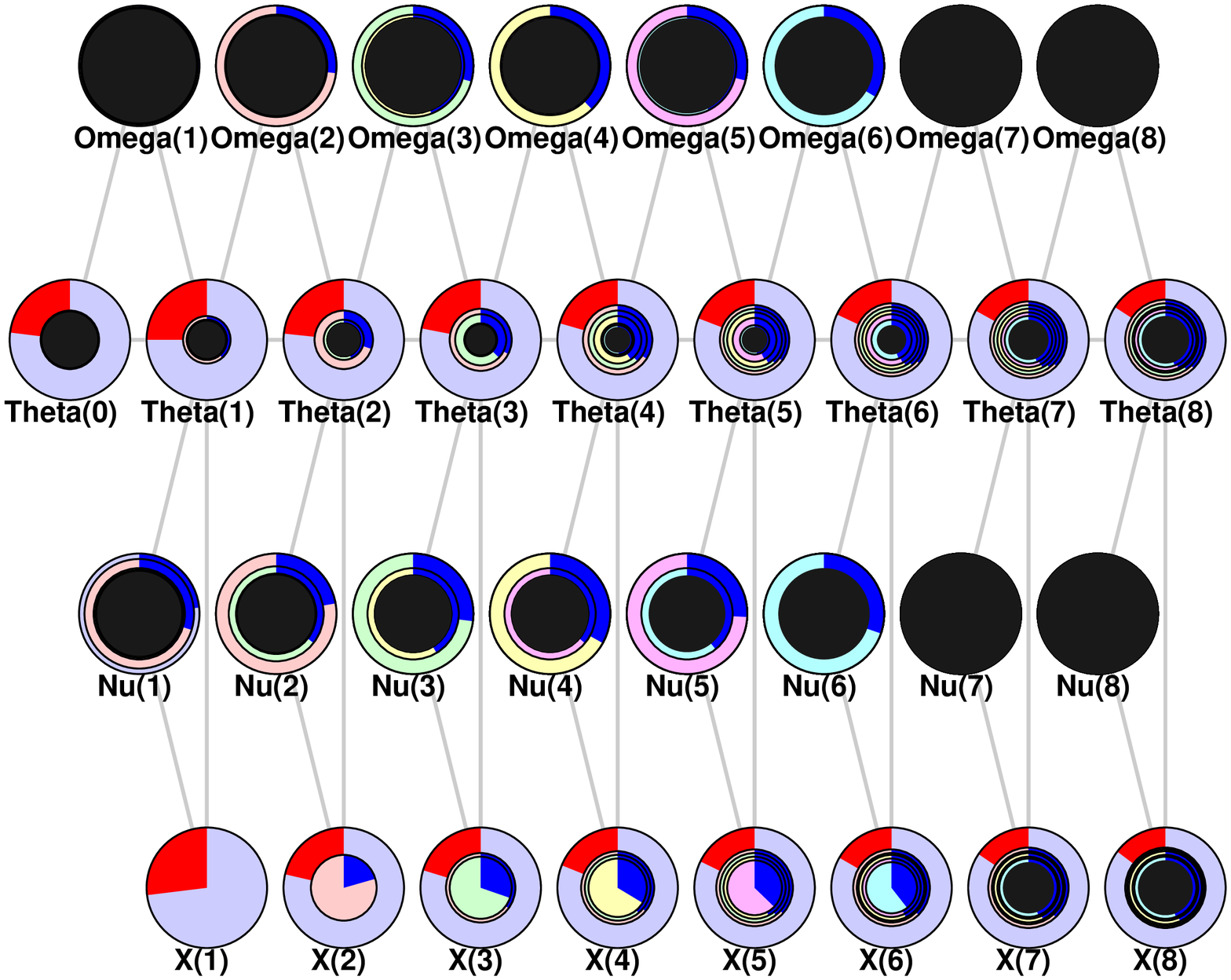,width=7in}

\noindent\epsfig{file=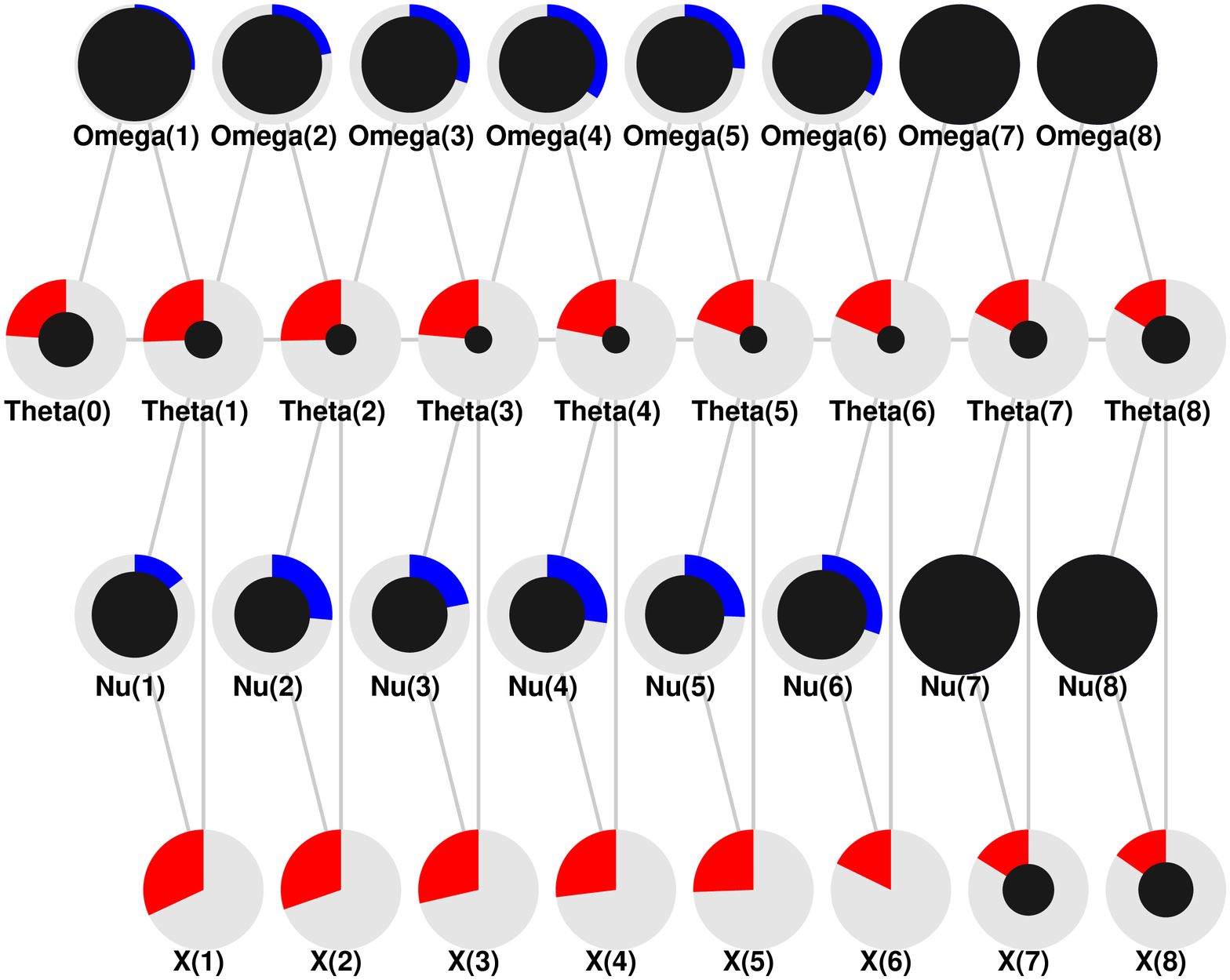,width=7in}

The file included as part of the distribution then goes on to extract
the adjusted expectations of the residuals, and plot them on a line
graph. See the example source file for more details.

\section{Important note/Disclaimer}

It is important to note that \blin\ is a rapidly developing prototype
system, and does contain many bugs. You should not rely on \blin\
producing correct output, and should verify parts of calculations as
far as possible, using alternative software, such as \bd\ (see
\cite{wgbd}). The author accepts no liability whatsoever regarding the
use of \blin, errors or losses arising from the use of \blin,
\emph{etc}. Feel free to check the source code, correct it, and email
the corrections to the author.

\section{Command reference}
This section lists all global variables, functions, object prototypes
and methods defined by the \blin\ system. On-line help is
available. To obtain help on a global function, \emph{eg.}
\verb$create-$ \verb$moral-$\verb$node$
use the expression 
\verb$(help 'create-moral-node)$
 To obtain help for a method,
\emph{eg.} the \verb$:observe$ method of the \verb$moral-node-proto$
object, use the expression \\
\verb$(send moral-node-proto :help :observe)$.

\subsection{Global variables}

\begin{center}
\begin{tabular}{|c|p{11cm}|}\hline
Variable & Description \\ \hline
\texttt{*tree-nodes*} &
A list of symbols representing instances of
tree-node objects created using the \texttt{create-tree-node} global
function.\\ 
\texttt{*moral-nodes*} &
A list of symbols representing instances of moral-node objects created
using the \texttt{create-moral-node} global function. \\ \hline
\end{tabular}
\end{center}

\subsection{Global functions}

\begin{center}
\begin{tabular}{|c|p{11cm}|}\hline
Function & Description \\ \hline
\texttt{create-tree-node} &
A function to create and initialise clique-tree objects.\\
\texttt{create-moral-node} &
A function to create and initialise moral graph objects.
\\
\texttt{create-moral-plot} &
A function to create a graphics window for illustrating and diagnosing
Bayes linear adjustments.
\\
\texttt{ginv} & Function to return the Moore-Penrose generalised
inverse of a real square symmetric matrix.\\ \hline
\end{tabular}
\end{center}

\subsection{Object prototypes}

\begin{center}
\begin{tabular}{|c|p{8cm}|}\hline
Object prototype & Description \\ \hline
\texttt{tree-node-proto} &
The prototype for objects representing clique-tree nodes. 
\\
\texttt{moral-node-proto} &
The prototype for objects representing moral graph nodes. 
\\
\texttt{moral-plot-proto} &
The prototype for a graphics window object for the displaying of
information relating to current Bayes linear adjustments. 
\\ 
\texttt{global-moral-plot-proto} &
A plot to summarise the partial adjustments shown on the
\texttt{moral-plot-proto} plots. \\ \hline
\end{tabular}
\end{center}

\subsubsection{\texttt{tree-node-proto} slots and methods}

\begin{center}
\begin{tabular}{|c|p{11cm}|}\hline
Slot & Description \\ \hline
\texttt{name} &
The name of the object.\\
\texttt{variables} &
List of random variables associated with this tree-node object.\\
\texttt{neighbours} &
List of neighbouring junction tree nodes.\\
\texttt{variance} &
Variance matrix associated with the variable list.\\
\texttt{expectation} &
Expectation vector associated with the variable list.\\
\texttt{var-d-inv} &
$\var{D}\inv$\\
\texttt{cov-d-self} &
$\cov{D}{\cdot}$\\
\texttt{obs-vars} &
$D$\\
\texttt{obs-d-ed} &
$d-\ex{D}$ \\
\texttt{location} &
On a $(0,1)\times(0,1)$ scale for plotting.\\ \hline
\end{tabular}
\end{center}

\begin{center}
\begin{tabular}{|c|p{11cm}|}\hline
Method & Description \\ \hline
\texttt{:absorb} & Absorb information from last \texttt{:observe}
ready for next observe.\\
\texttt{:aex} & Adjusted expectation.\\
\texttt{:avar} & Adjusted variance.\\
\texttt{:cov} & Current covariance.\\
\texttt{:ex} & Current expectation.\\
\texttt{:info} & Prints some information relating to the object.\\
\texttt{:location} & Accessor method.\\
\texttt{:observe} & Method to introduce data to the tree.\\
\texttt{:positions} & Variable positions.\\
\texttt{:propagate} & Method used to propagate information around the
tree.\\
\texttt{:remove-neighbour} & Remove a neighbour from the list.\\
\texttt{:resolution} & Partial resolution for the current
adjustment. \\
\texttt{:rvar} & Resolved variance matrix. \\
\texttt{:size-ratio} & Partial size-ratio for the current
adjustment.\\
\texttt{:transform} & Partial resolution transform for the current
adjustment.\\
\texttt{:var} & Current variance matrix. \\ \hline
\end{tabular}
\end{center}

\subsubsection{\texttt{moral-node-proto} slots and methods}

\begin{center}
\begin{tabular}{|c|p{11cm}|}\hline
Slot & Description \\ \hline
\texttt{name} &
Name of the object.\\
\texttt{variables} &
List of random variables associated with this moral node object. \\
\texttt{tree-node} & Name of a tree-node which contains all of the
variables in this moral node.\\
\texttt{print-name} & 
A string for plotting purposes.\\
\texttt{neighbours} &
A list of neighbouring moral graph nodes.\\
\texttt{location} &
On a $(0,1)\times(0,1)$ scale for plotting.\\
\verb$var_b_inv$ &
Inverse of the \emph{a priori} variance matrix for the variables
represented by this node.\\
\verb$ex_b$ &
The prior expectation vector for this node. \\ 
\texttt{resolutions} & List of resolutions for the partial
adjustments. \\
\texttt{size-ratios} & List of partial size-ratios for the
adjustments.\\
\texttt{global-size-ratios} & List of global size-ratios for the
adjustments.\\ \hline
\end{tabular}
\end{center}

\begin{center}
\begin{tabular}{|c|p{11cm}|}\hline
Method & Description \\ \hline
\texttt{:absorb} & Absorb info ready for next \texttt{:observe}.\\
\texttt{:aex} & Adjusted expectation vector.\\
\texttt{:avar} & Adjusted variance matrix.\\
\texttt{:bearing} & Bearing vector.\\
\texttt{:ex} & Current expectation vector.\\
\texttt{:info} & Prints some info about the object.\\
\texttt{:location} & Accessor method.\\
\texttt{:observe} & Introduce data into the graph.\\
\texttt{:remove-neighbour} & Remove a node from the neighbour list.\\
\texttt{:resolution} & Partial resolution wrt \emph{a priori}
structure.\\
\texttt{:rvar} & Partial resolved variance matrix.\\
\texttt{:size-ratio} & Partial size-ratio.\\
\texttt{:global-size-ratio} & Global size-ratio.\\
\texttt{:transform} & Partial resolution transform.\\
\texttt{:tree-node} & Accessor method.\\
\texttt{:var} & Current variance matrix.\\ \hline
\end{tabular}
\end{center}

\subsubsection{\texttt{moral-plot-proto} slots and methods}

This object inherits all slots and methods from
\texttt{graph-window-proto}, but also has the following.

\begin{center}
\begin{tabular}{|c|p{11cm}|}\hline
Slot & Description \\ \hline
\texttt{nodes} & List of nodes to be plotted.\\
\texttt{real-size} & Window size.\\
\texttt{radius} & Node radius (scaled).\\
\texttt{diagnostics} & Flag for diagnostic plotting. \\
\texttt{node-labels} & Flag for node label printing. \\
\texttt{outlines} & Flag for node outline printing. \\ \hline
\end{tabular}
\end{center}

\begin{center}
\begin{tabular}{|c|p{11cm}|}\hline
Method & Description \\ \hline
\texttt{:plot-arcs} & Draw the arcs associated with a given node.\\
\texttt{:plot-node} & Draw the given node, and all its shadings.\\
\texttt{:r-to-s} & Take ``real'' (screen) coords to scaled coords.\\
\texttt{:record} & Record current adjustment information for inclusion
in the plot.\\
\texttt{:redraw} & Guess!\\
\texttt{:resize} & Recalculate scale parameters.\\
\texttt{:s-to-r} & Scaled to real coord transform.\\
\texttt{:sr-map} & Function which maps $(0,\infty)$ to $(-1,1)$
monotonically, mapping $1$ to $0$. This is the function used to
transform size ratios for red and blue diagnostic shadings.\\
\texttt{:diagnostics} & Set and unset diagnostics plotting. \\
\texttt{:node-labels} & Set and unset node label printing. \\
\texttt{:outlines} & Set and unset node label printing.\\ \hline
\end{tabular}
\end{center}

\subsubsection{\texttt{global-moral-plot-proto} slots and methods}

This object inherits all slots and methods from
\texttt{moral-plot-proto} and has no others.

\newpage
\bibliography{bayeslin,djw}

\begin{thebibliography}{10}

\bibitem{fgcross}
M.~Farrow and M.~Goldstein.
\newblock {B}ayes linear methods for grouped multivariate repeated measurement
  studies with application to crossover trials.
\newblock {\em Biometrika}, 80(1):39--59, 1993.

\bibitem{mgrevprev}
M.~Goldstein.
\newblock Revising previsions: a geometric interpretation.
\newblock {\em J. R. Statist. Soc.}, B:43:105--130, 1981.

\bibitem{mgrevexch}
M.~Goldstein.
\newblock Revising exchangeable beliefs: subjectivist foundations for the
  inductive argument.
\newblock In P.~Freeman and A.F.M. Smith, editors, {\em Aspects of
  {U}ncertainty: A Tribute to D. V. Lindley}. Wiley, 1994.

\bibitem{blmadj}
M.~Goldstein.
\newblock {B}ayes linear methods {I} - {A}djusting beliefs: concepts and
  properties.
\newblock Technical Report 1995/1, Department of Mathematical Sciences,
  University of Durham, 1995.

\bibitem{mgpriorinf}
M.~Goldstein.
\newblock Prior inferences for posterior judgements.
\newblock In M.~L.~D. Chiara et~al., editors, {\em Structures and norms in
  science}. Pordrecht Kluwer, 1997.

\bibitem{gwblincomp}
M.~Goldstein and D.~A. Wooff.
\newblock Bayes linear computation: concepts, implementation and programming
  environment.
\newblock {\em Statistics and Computing}, 5:327--341, 1995.

\bibitem{pearlbook}
J.~Pearl.
\newblock {\em Probabilistic reasoning in intelligent systems}.
\newblock Morgan Kaufmann, 1988.

\bibitem{lispstat}
L.~Tierney.
\newblock {\em LISP-STAT: An object oriented environment for statistical
  computing and dynamic graphics}.
\newblock Wiley, 1990.

\bibitem{djwgdlm}
D.~J. Wilkinson and M.~Goldstein.
\newblock Bayes linear covariance matrix adjustment for multivariate dynamic
  linear models.
\newblock {\em U. Ncle. Stats. Tech. report STA97,12},
  \texttt{http://www.ncl.ac.uk/\~{}nstat/preprints/}, 1997.

\bibitem{wgbd}
D.~A. Wooff and M.~Goldstein.
\newblock {[B/D]} --- {B}eliefs adjusted by {D}ata: {B}ayes linear methods
  programming language.
\newblock {\em Internet site \texttt{http://fourier.dur.ac.uk:8000/stats/bd/}},
  1995.

\end{thebibliography}
\bibliographystyle{plain}

\end{document}